# Stress controlled magnetic properties of Cobalt nanowires

**Jaivardhan Sinha and S. S. Banerjee***
Department of Physics, Indian Institute of Technology, Kanpur-208016, U. P., India.

*E-mail: satyajit@iitk.ac.in

**Abstract:** We investigate the magnetic properties of a composite comprising of ferromagnetic Cobalt nanowires embedded in nanoporous anodized alumina template. We observe unusual increase in, the saturation magnetization and the coercive field, of the nanowires below 100 K. We also report the appearance of an unusual exchange bias effect in nanowires below 100 K. We argue our results can be understood on the basis of a competition between different magnetic energy scales induced by significant stresses acting on the nanowires at low temperatures. The composite behaves as an effective medium in which the magnetic anisotropy of nanowires can be conveniently controlled via stress on the nanowires.

Properties of nanostructured materials are sensitive to changes in morphology, composition and strain. From applications point of view, the ability to control strain in nanostructures is an attractive proposition as it provides a convenient route to manipulating their magnetic and transport properties. In recent times, work on coherently strained core-shell structured nanowires of semiconducting materials has shown potential for application in diverse areas ranging from electronics, photonics to biology [1, 2, 3]. Motivated by a similar goal, in the present work we have investigated a composite with nanowires in which stress on the nanowires can be effectively varied to control the magnetic properties of the nanowires. Amongst magnetic materials we investigate the magnetic properties of Co nanowires. In recent times, magnetic properties of nanostructures of Co, Ni and Fe have been extensively investigated [4,5,6,7,8,9,10,11,12,13,14,15,16,17]. In magnetic materials usually two dominant competing interactions determine the final orientation of magnetization, viz., (i) the magneto-crystalline anisotropy, which due to strong spin-orbit interactions prefers to align the moments along a specific crystallographic direction, and (ii) the shape anisotropy, wherein depending on the crystal shape, non uniform demagnetizing field orients the magnetization along a specific geometrical direction of the material. Unlike Ni [16,18], in Co nanowires the relatively large magneto-crystalline anisotropy of Co competes with the enhanced shape anisotropy in a nanowire to determine the final orientation of magnetization in the nanowire. Due to this the magnetic properties of Co nanowires are different from those of Ni or Fe. In Co nanowires two different orientation of magnetization have been reported, viz., oriented either perpendicular or parallel to the nanowire long axis [18,19], depending on the diameter of the wire being above or below 50 nm respectively. In some systems the magnetic anisotropy is governed by another strong effect, viz., the Exchange Bias (EB) effect [20].The EB effect which is usually found across an Ferromagnet (FM) – Antiferromagnetic (AFM) interface, produces strong pinning of the magnetic anisotropy (orientation) along a specific direction due to exchange interaction across the FM – AFM interface. Due to the exchange interaction across the interface, the pinning of FM moments by the AFM layer makes it is difficult to flip the moments away from the preferred orientation as compared to returning back the moments to their original preferred orientation by cycling the externally applied magnetic field across the coercive field of the material. This produces a horizontal shift in the hysteresis loop [20] below the AFM ordering temperature, which is a characteristic signature of the EB effect. As the EB effect aids in stabilizing the magnetic anisotropy along a particular direction, therefore it is a useful property to engineer and control for the magnetic data storage and recording industry [21,22]. From applications point of view it would be a lucrative proposition to investigate the possibility of a mechanism to physically control the EB effect in a magnetic structure rather than chemically produce AFM layer on a FM. In this letter we investigate the temperature dependent magnetic properties of a composite comprising of Co nanowires encapsulated in a nanoporous Anodized Alumina (AAO) template. Our investigations reveal unusual enhancement in the saturation magnetization and the coercive field of Co nanowires as the temperature of the matrix is reduced below 100 K. Furthermore, as the temperature of composite matrix is lowered below 100 K a horizontal shift appears in the hysteresis loop indicating a temperature dependent EB effect. Analysis of our observations suggests temperature induced stress on the nanowires as being responsible for altering the magnetic configuration and properties of the Co nanowires at low temperatures.

Home grown nanoporous AAO template was produced using standard electrochemical routes [23] and used for growing the nanowires. Figure 1(a) is a Scanning Electron Microscope (SEM) image of our AAO template surface which has pores with uniform pore size, with an average diameter of 60 nm ± 5 nm. The pores are well ordered with a center to center distance of $D \sim 115$ nm and average pore density of 50-100 /$\mu m^2$. Inset Fig 1(b) shows a fast Fourier transform (FFT) of the nanoporous AAO template shown in Fig.1(a). One observes six slightly diffused diffraction spots in the FFT image (cf. Fig. 1(b)) indicating reasonably well ordered hexagonal array of nanopores in the AAO template. The inset Fig. 1(c) shows the SEM image of the cross-section of the AAO template indicating highly oriented parallel pores extending throughout the entire thickness of the template. By sputtering Co onto the AAO template under inert atmosphere of Ar gas at a pressure of $2.2\times10^{-2}$ mbar with the template held at 325 °C, we obtained nanowires of Co confined within the AAO nanopores. Figure 1(d) shows the Transmission Electron Microscope (TEM) (Model No FEI Technai 20 U Twin Transmission Electron Microscope) image of the Co nanowires (dark contrast) inside the AAO nanopores (bright contrast are obtained from the walls of the AAO nanopores). The average width of the Co-wires confined inside the AAO pore is, $d \sim$ 40 - 50 nm. The presence of Co in the (dark) channels was confirmed by spot Energy dispersive X-ray (EDAX) spectroscopy measurements. A few of the Co nanowires we found to extend up to almost 5µm in length inside the nanopores. After dissolving the AAO template, the crystal structure of the Co nanowire was characterized by capturing Selected Area Diffraction Pattern (SADP) image using the TEM. In Fig. 1(e), the dashed (yellow) circles are the locus of diffraction spots from the HCP crystallographic planes of Co (under ambient conditions Co is usually found in HCP phase with $c/a \sim 1.62$ [24])). In Fig. 1(e) we also observe a set of bright spots located on the (red) dotted circle which do not correspond to diffraction from the HCP structure. The diffraction spots on the (yellow) dashed circle were indexed to the (100), (103), planes of the HCP (pcpdf-05-0727) structure while the non – HCP diffraction spots (on the (red) dotted circles) matched with the (200), (220) planes of the FCC structure (pcpdf-15-0806) of Co. Earlier structural investigation of Co nanowires grown via different techniques have also revealed the presence of the HCP and FCC phases [6,25,26].

Magnetization (*M*) response of the matrix of Co nanowires embedded in the Alumina template (size of 2.5mm×2mm×50µm, 50µm is the average thickness of AAO template) was measured using a commercial Superconducting quantum interference device magnetometer (SQUID, from Quantum Design). Unlike most measurements which are on free standing nanostructures, we measure the response of the entire composite without dissolving the AAO template. The AAO template not only ensures that the Co nanowires are oriented parallel to each other but also helps to maintain a reasonable separation between the nanowires resulting in weak inter-wire interaction. Figure 2 (a) and the inset panel (b) display the isothermal magnetization hysteresis loops ($M(H)$). All the $M(H)$ loops are recorded in the field range ±70 kOe (cf. inset Fig.3), however on order to discern features in the hysteresis loop, the data in Fig. 2 is plotted between ±35 kOe. Two different orientations of the Co nanowires long axis w.r.t. the magnetic field is achieved by aligning the AAO template surface to be either perpendicular (⊥) to or parallel (//) to the applied magnetic field ($H$). In either orientation the magnetization response along the applied field direction (which is always vertical) is measured. Data shown in Fig. 2 is obtained after subtracting the isothermal paramagnetic $M(H)$ background response for the bare AAO template (of same dimensions) without Co. From Fig.2 (a), note that for 100 K < $T$ < 300 K, the $M_\perp$ ($H$) hysteresis loops (viz., $H$ ⊥ to the wire long axis) is squarer in shape as compared to the shape of $M_{//}$ ($H$) loops ($H$ // to the wire long axis, cf. Fig. 2(b)). The square shape of the $M_\perp$ ($H$) loop implies that the magnetization easy axis is oriented perpendicular to the wire long axis. This inference is consistent with reports that the crystallographic c-axis of the HCP phase of Co which is also the easy magnetization axis, is aligned ⊥ to the wire long axis [18,19,26] in Co-nanowires with diameter in the range of 50 nm. The degree to which the shape of the hysteresis loop is square is measured via the ratio of the remnant magnetization ($M_{re}$) to the saturation magnetization ($M_s$). Ideally an $M_{re}/M_s$ = 1, indicates a perfect square shape of the $M(H)$ hysteresis loop. Figure 3 shows that for $H$ ⊥ to the wire long axis, in the temperature range of 100 K < $T$ < 300 K, $M_{re}/M_s$ = 0.7 indicating a nearly square shape of hysteresis loop. At $T$ < 100 K, Fig.3 shows the hysteresis loop significantly deviates more from the square shape, viz., ($M_{re}/M_s$) decrease to far below 0.7 at low $T$ (< 100 K), indicating that the orientation of $M$ perpendicular to the wire long axis is no longer preferred at low $T$. In Fig.3 the observation that $M_{re}/M_s$ ratio decreases to well below 1, suggests changes in the magnetization easy axis with lowering of T. From the similarity of the shape of the magnetization loops below 100 K in either orientation in Fig.2 suggests that the anisotropic magnetization response of the Co nanowires decreases below 100 K.

With the changes in the easy axis of magnetization at low $T$, we also find a significant increase in $M_s$ as $T$ is lowered below 100 K. In Fig. 2(a)), the increase in $M_s$ is by almost 160 – 170 % at 3 K compared to its value at 300K. The observed changes in the shape of the loop viz., deviation in square shape below 100 K along with the increase in $M_s$ below 100 K, implies, above 100 K moments which were earlier not aligned perpendicular to the wire axis begin to align along $H$, below 100 K, which leads to the enhancement in $M_s$. Similar increase in $M_s$ is also seen for $H$ applied // to wire long axis (Fig. 2(b)). It is indeed an unusual observation that the moments appear to be gaining freedom to align along the field direction at low $T$ (< 100 Oe) in contrast to behavior at high $T$ where anisotropy effects predominate the magnetization response. In systems like Co with large magneto-crystalline anisotropy energy, the moments are usually constrained to orient along a specific crystallographic orientation which maybe undisturbed even by the application of large fields, therefore the observation of freeing up of moments to orient along the field hints towards a reduction in the magneto-crystalline anisotropy of Co at low $T$. We will dwell on this issue in a later section.

Apart from the changes discussed in $M_s$ below 100 K, we also find significant asymmetry appearing in the magnetization hysteresis loop ($H \perp$ to wire) below 100 K. Figure 4 main panel and Fig.4(b) demonstrates the presence of a horizontal asymmetry(shift) in the $M(H)$ hysteresis loops at $T = 3K$ and 40K. The curves shown in Fig. 4 main panel and in inset Fig. 4(b) are obtained by horizontally shifting the reverse leg (red) of $M(H)$ curve by $2H_c$ along field axis ($H_c$ is the temperature dependent coercive field which can be determined from the $M(H)$ loop (cf. Fig.2). Unlike a conventional FM material, the presence of a field gap between the forward (black) and the shifted reverse (red) leg of the *M(H)* curve is clearly established. In Fig. 4(a) we have plotted half of this field gap denoted as $H_{EX}$ (as described above) as a function of *T* (data points are the black square). Notice that $H_{EX}$ decreases with increasing *T* (cf. Fig.4(a)) and becomes undetectably small beyond 100 K. The observed horizontal asymmetry in hysteresis loop associated with $H_{EX}(T)$ is the manifestation of EB effect in Co nanocomposite [20-22]. Below 100 K, $H_{EX}$ increases by almost one order of magnitude. Note, all of the above magnetization response is reproducible over repeated cycling of temperature of the composite between 300 K down to 3 K. Note that in our experiments on the EB effect it was not possible to perform cooling field experiments, viz., experiments which attempt to probe the origin of the EB effect by warming up the sample to above the blocking temperature and then cooling it in a field and measuring the dependence of $H_{EX}$ on the cooling field [20,34]. The blocking temperature ($T_B$) is a temperature associated with thermal activation of uniformly magnetized domains over anisotropy energy barriers. Above $T_B$ the hysteresis in magnetization disappears [27]. In our system we observe the persistence of magnetization hysteresis in our Co nanowires upto 350 K (cf. Fig.2) suggesting a high blocking temperature which is beyond 350 K. To avoid to the possibility of damaging the composite and the limitation of reaching a high temperature range in our SQUID magnetometer it was not possible to perform cooling field dependence studies of $H_{EX}$ [20,34] on our nanowires.

At first glance it may seem natural to attribute the EB effect observed in our Co nanocomposite to the formation of CoO on the Co nanowire surface. The AFM CoO layer on the nanowire surface exchange coupled to the FM Co nanowire core would be a natural source of EB effect. TEM studies have routinely been used to establish either the presence or absence of a shell of CoO on a core of Co [28,29]. By grinding the composite we obtain a thin layer of Co (see the (red) dotted circle in Fig. 5(a)) attached to the AAO substrate. This thin Co layer attached to the AAO template is representative of a cross-section of our Co nanowire embedded in the AAO matrix. In the lower inset of Fig. 5(a), the SADP from this thin Co layer shows diffraction spots corresponding to the (200) FCC lattice of Co. In this lower inset of Fig. 5(a) we have also drawn a black circle to identify the location where one expects to observe diffraction spots due to (111) plane of CoO (pcpdf-65-2902). We find no diffraction spots located on the dark circle drawn. A high resolution TEM image of the Co nanostructure shown in the upper inset of Fig. 5(a) indicates lattice planes with d = 1.77 A°, which correspond to the (200) FCC lattice plane of Co which is consistent with the information in SADP. Here too there is no evidence of CoO lattice planes. To supplement this information we refer to Fig. 5(b) which shows the spot EDAX spectrum from the Co piece in Fig. 5(a), where the Co and Al peaks (due to slight contribution from AAO) can be identified. It is worth mentioning here that in the EDAX spectrum too we do not observe any significant Oxygen peak which should occur around 0.5 keV. These indicate that in our Co nanostructure there is no significant presence of CoO coating on the surface of the nanowire. Earlier studies on Co nanowires grown by electro-deposition in nanoporous AAO templates as well as in nanoporous polycarbonate membranes also report the absence of oxidation on the surface of the nanowires [30], which are similar to our findings. Furthermore recall that in conventional Co-CoO nanostructures where there is a significant amount of CoO present on the nanostructure, the magnitude of the $H_{EX}$ in such structures are reported to be ~ 1 kOe at low temperatures of 50 K [31]. For our composite, $H_{EX}$ is much smaller than 1 kOe and is of the order of 100's Oe at 3 K. Based on all of the above evidences, we summarize that though CoO maybe present in insignificantly small amount, it difficult to attribute the EB effect entirely to its presence. Current literature on the EB effect indicates that EB need not only be produced when an FM layer is exchange coupled to an AFM. In fact EB has been reported in a FM in contact with a material with disordered spin configuration like in a spin glass [32]. Recent studies also suggest a scenario where freezing of uncompensated spins leads to a disordered spin configuration at the surface of the FM. Presence of a significant exchange interaction across the interface separating a FM core from a shell with disordered spin configuration has also been proposed as a scenario leading to EB effect [33,34]. We investigate the possibility of producing a disordered spin configuration due to thermal fluctuations effects as a possible source of the EB effect.

Figure 6 shows (with solid data points) the temperature dependence of the coercive field $H_c(T)$, measured for field perpendicular to the wire long axis determined from the $M(H)$ loops in Fig. 2 (see the location of $H_c(T)$ marked in Fig. 2). In Fig.6 we find that the slope of the $H_c(T)$ curve in the vicinity of 100 K changes quite sharply in comparison to that in any other temperature regime (see the solid squares in Fig. 6). Physically $H_c(T)$ corresponds to the field required for magnetization reversal. Effects of thermal fluctuations are reflected in the temperature dependence of $H_c(T)$. Thermal fluctuations would lead to thermal activation of moments across the magnetic anisotropy energy barrier, which in turn would be reflected in the temperature dependence of the $H_c$. The thermally activated monotonic decrease in $H_c(T)$ with $T$ as [35] is given as, $H_c(T) = H_{c0} \frac{M_s(T)}{M_{s0}} \left[ 1 - \left( \frac{25 k_B T M_{s0}^2}{E_{00} M_s^2(T)} \right)^{1/m} \right]$, where $M_s(T)$ is the temperature dependence of the saturation magnetization of our Co – nanowires (see inset Fig. 6(b)), $M_{s0} = 9.75 \times 10^{-5}$ emu is the extrapolated value of $M_s(T)$ at $T = 0$ K, m = 3/2 and $E_{00} \sim 2.5 \times 10^{-11}$ erg is the value [35] of the shape anisotropy energy barrier at $T = 0$ K and $H = 0$ Oe for our Co-nanowire diameter of 50 nm. Figure 6(a) compares the experimentally measured $H_c(T)$ (solid data points) for our 50 nm Co nanowires with the $H_c(T)$ behavior expected from thermal fluctuations alone in free standing Co nanowires modeled by the above equation (see the dashed curve in Fig. 6(a)). From Fig.6 the actual $H_c(T)$ behavior is found to significantly differ from the thermally activated $H_c(T)$ behavior given by the above equation, where there is also no similarity in the curvature of the actual $H_c(T)$ data with the behavior expected from the above expression. It is known that presence of EB effect causes a significant enhancement in the anisotropy energy barrier [20]. Our observation of onset of EB effect in Co nanowires below 100 K corresponds well to the observation of a sudden increase in $H_c(T)$ below 100 K. However if origin of the EB effect in our nanowires were due to freezing of spins caused by reduction in thermal fluctuations below 100 K, then we should have observed a closer match of the $H_c(T)$ behavior with the above equation representing thermally activated behavior affecting $H_c(T)$, which is absent. Based on this we argue against the possibility of generating a disordered spins configuration on the surface of the nanowire due to thermal fluctuations, and is not the source of EB effect in our Co nanowires.

Absence of, significant amounts of Co-CoO interface or thermal fluctuation effects leading to freezing a disordered spin configuration to explain the observed EB effect in our Co nanowires, we investigate the possibility of an alternative mechanism. As the Co nanowires are encapsulated in a matrix, we investigate the role of thermal stresses in the composite as the temperature is lowered. A simple heuristic calculation based on procedure outlined in Ref. [36], shows a

$$strain(\varepsilon) = \int_{T_i}^{T_f} \alpha_{Co} dT \sim |\alpha_{Co}(T_f - T_i)| \sim 4\times10^{-3}$$

on the Co nanowire as it is cooled from $T_i = 300$ K to $T_f = 3$ K, where thermal expansion coefficient of Alumina and Co, $\alpha_{alumina} = 6 \times 10^{-6}$ K$^{-1}$ and $\alpha_{Co} = 1.3 \times 10^{-5}$ K$^{-1}$ (and neglecting their $T$ dependence) were used [5]. Using the Young's modulus ($E_{eff}$) of Co ~ $1.6 \times 10^{11}$ Pa [36] we estimate the stress ($\sigma$) on the wires as ~ $\varepsilon E_{eff}$ ~ 1 GPa. The above estimate we believe is only a lower estimate of the stress developing in the composite at low $T$. It is plausible that large stress on the nanowire surface decreases in the c/a ratio of the HCP lattice whose crystallographic c-axis are aligned almost perpendicular to wire long axis. It is known for Co, that the c/a ratio ~ 1.62 for the HCP phase of Co under ambient conditions and it decreases significantly with increase in pressure on the lattice[24]. The lowering of the c/a ratio of the HCP phase on the nanowire surface in-turn is expected to trigger a decrease in the magneto-crystalline anisotropy due to an enhancement in the symmetry of the compressed

HCP phase (as the c/a approaches closer to 1). Consequently it is expected that the contribution from shape anisotropy should effectively increase at low T as the magneto-crystalline anisotropy contribution decreases. Based on this expectation we investigate the balance between different magnetic anisotropy energies in the Co nanowire, to uncover indirect evidence of strain on the nanowires.

Figure 7 shows the temperature variation of the ratio of magneto-crystalline anisotropy energy density estimated using the expression [36] $K_{eff} = \frac{1}{2}\mu_0 M_s \left(H_\perp^S - H_{//}^S\right)$ to the shape anisotropy energy density $K_{shape} = \frac{\mu_0 M_s^2 \Delta N}{2}$, where $H_\perp^S$ and $H_{//}^S$ are the saturation field for applied field direction $\perp$ and $//$ to the wires (determined from Fig. 2), $M_s$ is the saturation magnetization and the demagnetization factor $\Delta N$ (~ 0.7 for a packing fraction($P$) of 0.05 for our composite, see Ref. [7]). Using the above expression at 300 K we have estimated for Co $K_{eff}$ ~ 6.3 x $10^6$ erg/cc ( $\approx$ $K_{shape}$), which compares well with values reported in literature [37]. In Fig. 7 it appears that just below 100 K the ratio begins to fluctuate substantially between extremes of nearly 1.5 and close to zero. The observed fluctuation in Fig. 7 suggests a changing balance between the $K_{eff}$ and $K_{shape}$ energy scales below 100 K, with the $K_{eff}$ decreasing significantly relative to $K_{shape}$ at low T. As discussed earlier, consistent with our expectations due to stress induced distortions in the crystallographic c/a ratio at low T, we find that the contribution of $K_{eff}$ decreases significantly relative to $K_{shape}$ at low T (< 100 K).

The lowering of magneto-crystalline anisotropy ($K_{eff}$) at low $T$ (< 100 K), would allow moments to begin orienting themselves with the applied field direction which were originally (at T > 100 K, due to higher $K_{eff}$ at these T) constrained to orient along a specific crystallographic orientation which is distinct from the applied field direction. This is the source of the unusual observation of an increase in $M_s$ at low $T$ (cf. Fig. 2). In our nanowires it is unlikely that all the HCP crystallites have their c-axis, which is also the easy axis of magnetization, oriented perpendicular to the nanowire. Crystallites which have their c-axis oriented in direction other than the H direction due to strong $K_{eff}$, do not contribute significantly to $M_s$ at high T (> 100). At high T (> 100 K) a large applied field (more than 70 kOe, which is our present limit of measurement) is required to overcome the magneto-crystalline anisotropy and align all the moments along the applied field direction (including those moments which are constrained by magneto-crystalline anisotropy ($K_{eff}$) to be aligned along the crystallographic orientation different from the H direction). At low T (< 100 K) due to high stress induced reduction in magneto-crystalline anisotropy ($K_{eff}$) allows for moments irrespective of orientation w.r.t the easy axis of the nanowire to begin aligning along the applied field direction. Therefore moments which had contributed little to the net $M_s$ at high T (> 100 K) now begin contributing to $M_s$ at low T (< 100 K). It is tempting to suggest that compared to Ni or Fe, Co which had a highly anisotropic magnetization response at high T (> 100 K), at low T (<100 K) the Co nanocomposite begins to have a more isotropic magnetization response and appears to almost behave like Ni or Fe due to a stress induced reduction magneto-crystalline anisotropy, where shape anisotropy effects govern the magnetization response behavior.

At low T (< 100 K) the competition between the enhancing shape anisotropy contribution for the nanowires (with large aspect ratio) relative to reducing magneto-crystalline anisotropy contribution promotes a disordered configuration of moments on the nanowire surface. The nanowire core which is much less affected by thermally induced stress compared to the surface possess moments whose orientation are still predominantly governed by magneto-crystalline anisotropy. While the nanowire wire surface possess a disorder moment configuration. Such a heterogeneous magnetization configuration with a core of ordered FM moments and a shell of disordered moments, is similar to a core - shell like configuration which is known to promote the EB effect [34]. The onset of EB effect produces an enhancement in $H_c$ below 100 K (see Fig. 6(a)). To better understand the unusual temperature dependence of $H_c(T)$ in Fig. 6(a) one needs to understand better the magnetization anisotropy energy barrier height which is governed by stresses on the nanowire. We would like to reiterate that the composite is very robust and the reported unusual magnetic properties are reproduced over repeatedly cycling between 3K and 300 K repeated in intervals of several weeks.

In summary, we have investigated the temperature dependent magnetization behavior of Co nanowires embedded in a nanoporous alumina template. We find the magnetic anisotropy of the Co nanowires is effectively controlled and varies with the temperature of the composite. Reducing the temperature of the composite leads to enhanced saturation magnetization, high coercive field, and appearance of an unusual EB effect. We argue our composite provides a route for producing an EB effect in the nanowire by varying an external physical parameter like stress on the nanowire. By controlling the temperature of composite and hence the stress on the wires one creates an interface separating an ordered FM core with a shell of disordered spin configuration leading to stress controlled EB effect. We propose that our composite with Co nanowire could be a potential candidate for perpendicular magnetic recording. It appears that in the composite at low T at moderately low field one should be able to reorient the magnetization along the Co nanowire axis with relative ease. Thus, this composite has the potential to modulating the magnetization anisotropy by variation of temperature of the composite.

**Acknowledgements**: SSB thanks Department of Science and Technology India and Indian Institute of Technology Kanpur, India for funding support. We also thank Dr. Shyam Mohan and Mr. C. S. Srivatsan for their help during experiments. SSB would like to thank the Nanoscience center at IIT Kanpur for access to facilities like the TEM and also thanks Mr. Rajiv Sharma for his help in running the SQUID magnetometer at IIT Kanpur.

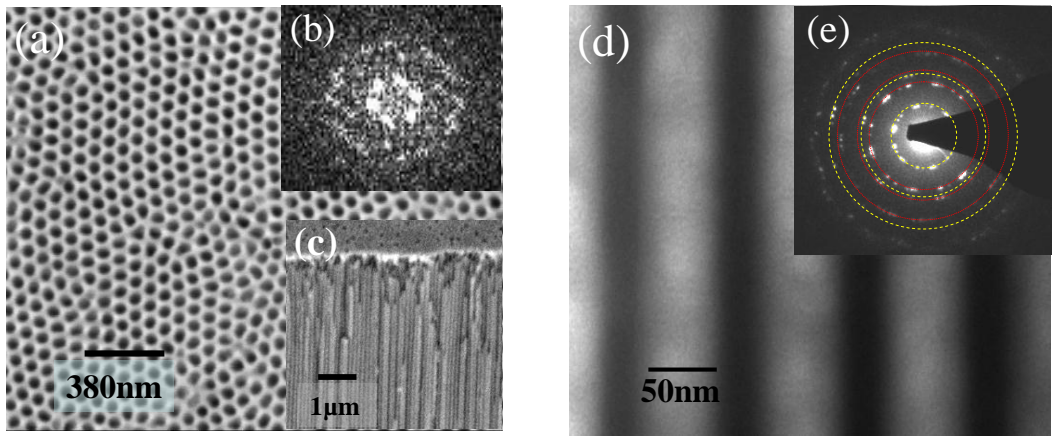

**Figure 1:** (a) Shows the SEM image of nanoporous alumina template. (b) FFT of image of the nanoporous alumina template shown in Fig. (a). (c) Shows the cross sectional view of the nanoporous alumina template. (d) TEM image of the Co nanowires confined in alumina matrix. (e) Selected area diffraction pattern of the Co nanowires (cf. text for details).

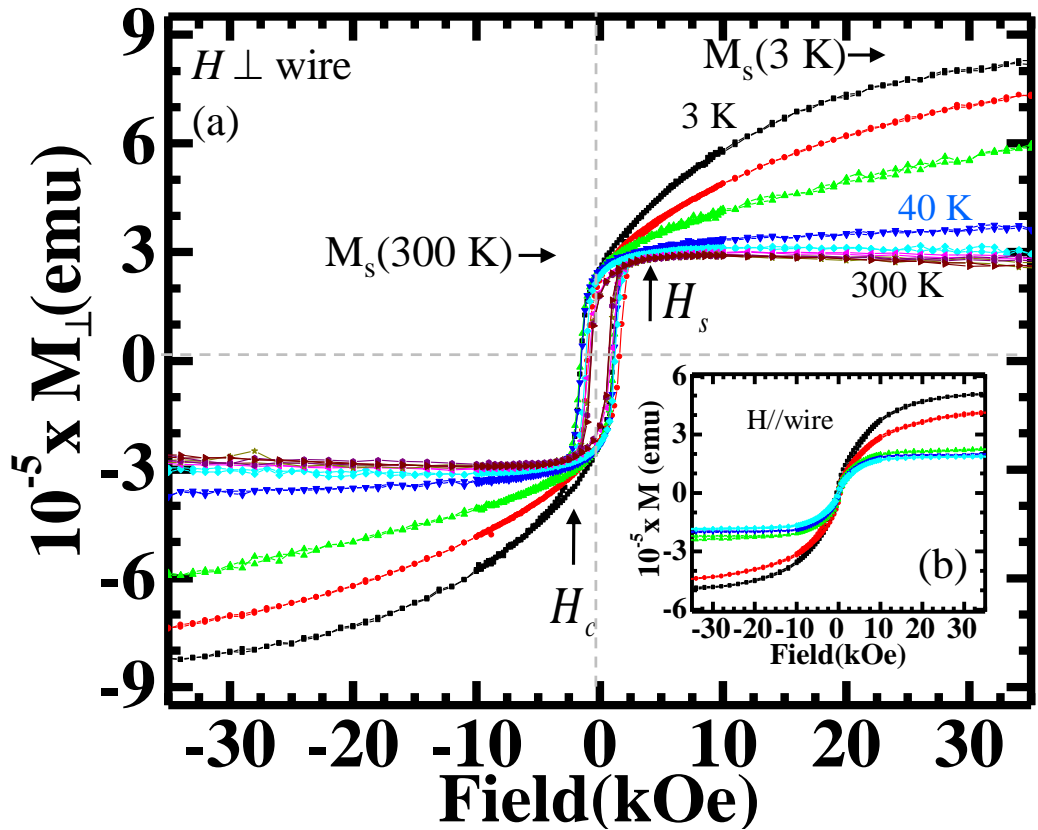

**Figure 2:** (a) Shows the magnetization hysteresis (*MH*) loop at different temperature from 300K to 3K for the applied field direction perpendicular to the nanowire. The legends correspond to 3K (■), 5K (●), 10K (▲), 40K (▼), 100K (♦), 150K (◄), 200K (★), 250K (●) and 300K (►) in the figure. Inset (b) shows the *MH* loops at different temperature values 3K (■), 5K (●), 100K (▲), 200K (▼) and 250K (♦) for applied field parallel to the nanowires.

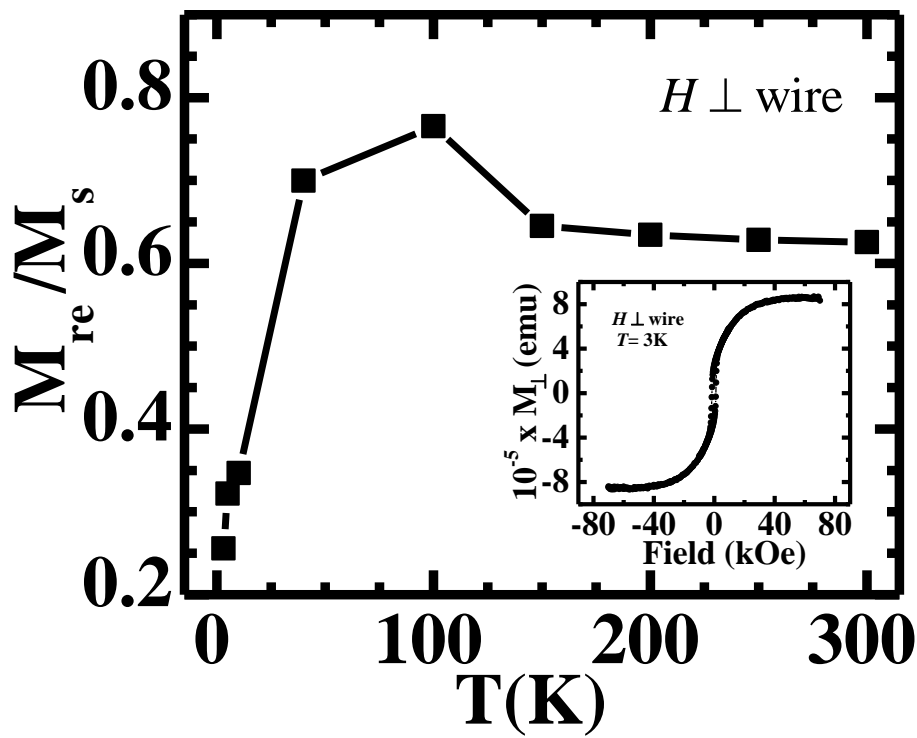

**Figure 3:** Shows the variation in remnance ratio (squareness) of the loop as a function of temperature. The inset shows the magnetization hysteresis loop at 3K measured upto ±70kOe for field applied perpendicular to the nanowire.

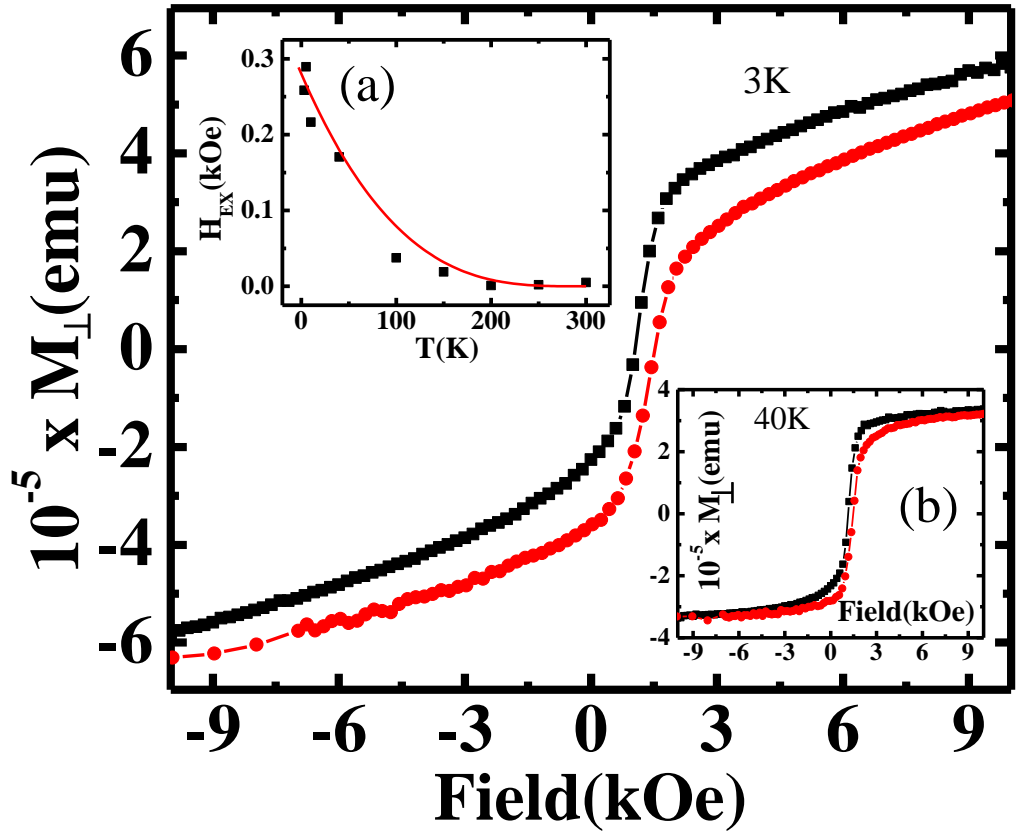

**Figure 4:** Main panel shows the horizontal shift (asymmetry) in hysteresis loop at 3K. Inset (a) shows the behavior of exchange bias field ($H_{EX}$) as a function of temperature. The solid red curve is a guide to the eye. Inset (b) shows the horizontal shift in hysteresis loop at 40K. (cf. text for details).

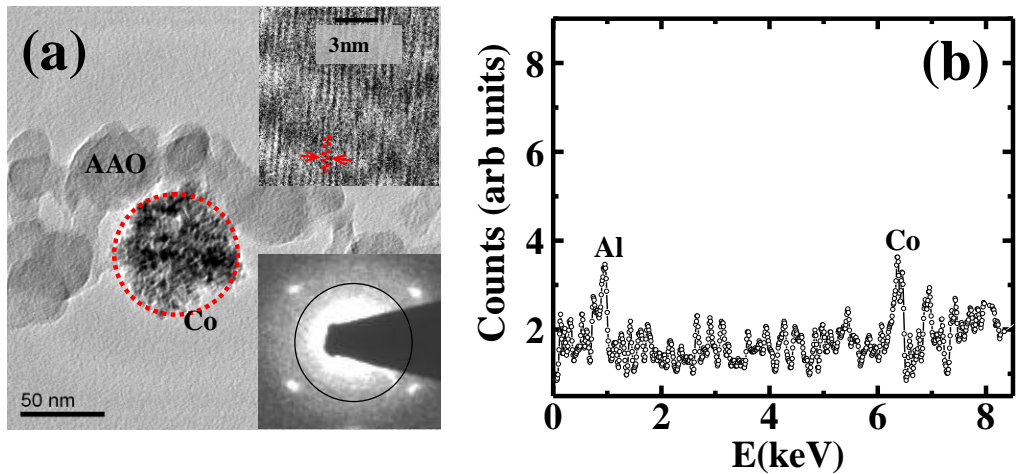

**Figure 5:** (a) Main panel shows a TEM image of a piece of Co attached to AAO template which is derived out of the nanostructure. Lower inset shows the SADP from the nanostructure. The circle drawn in this image shows the distance at which diffraction spots from CoO should have present (cf. text for details). Upper inset shows the high resolution TEM image showing the lattice planes corresponding to a non HCP structure (cf. text for details). (b) Shows the EDAX spectrum taken from the Co nanostructure shown in figure (a).

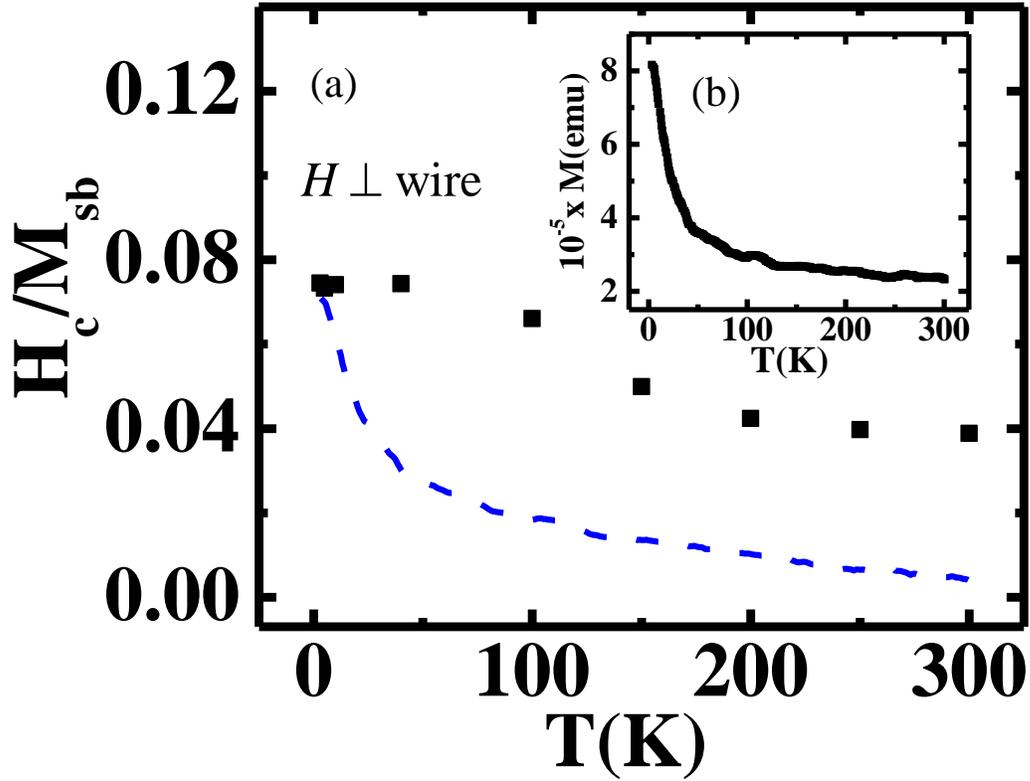

**Figure 6:** (a) Shows the variation of coercive field with temperature (black filled square) from magnetization hysteresis loop. Dashed (Blue) curve shows the calculated coercive field values using expression mentioned in text for different data points from *MT* data (cf. text for details). Inset (b) shows the variation of saturation magnetization as a function of temperature for an applied field of 70 kOe perpendicular to the wire.

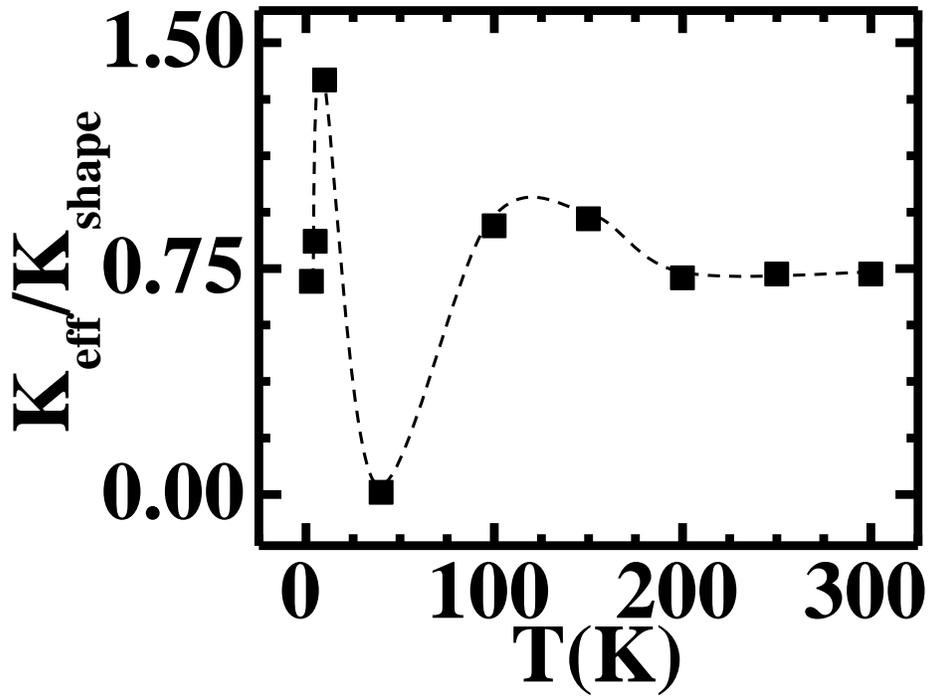

**Figure 7:** Variation of ratio of $K_{eff}$ to $K_{shape}$ as a function of temperature (cf. text for details).